
\font\titlefont = cmr10 scaled \magstep2
\magnification=\magstep1
\vsize=22truecm
\voffset=1.75truecm
\hsize=15truecm
\hoffset=0.95truecm
\baselineskip=15.5pt

\settabs 18 \columns

\def\b{\bigskip}

\def\ce{\centerline}

\def\no{\noindent}

\ce{\titlefont{Some Recent Developments in Sphalerons }}

\b
\ce{Bing-Lin Young}

\ce{Department of Physics and Astronomy }
\ce{ Iowa State University,}
\ce{ Ames, Iowa
 50011}
\b
\ce{\bf ABSTRACT}
\no We review briefly the sphaleron and list
some of its properties. We summarize some of the results in
models which have an extended scalar sector.  We also present our
work on models dealing with physics beyond the standard model.
We focus on the energy of the sphaleron which is important in determining
the rate of baryon number violation at the electroweak scale.
  \b
\no{\bf I. Introduction}

In the standard model baryon number and lepton number are not
conserved due to the existence of the anomaly, and non-trivial
vacua topology of the theory, as pointed out by
't Hooft[1].
Manton and Klinkhamer[2]
showed that in
the SU(2) gauge-Higgs theory
there exist non-contractible loops in the configuration
space which is composed of stationary, finite energy solutions of the
classical equations of motion. The highest-energy configuration on a
minimal energy path is a saddle point and it can be interpreted as the
minimal energy barrier separating the neighbouring vacua of different
Chern-Simon
numbers.
This is called a sphaleron[2]. The
minimal height of the barrier is the energy of the sphaleron which will
be denoted as $E_{sp}$.

This interpretation of the sphaleron led to its application to
electroweak baryogenesis[3].
In the high temperature environment of the early universe thermal
excitation becomes important, and the transition from one vacuum to another
can be effected by passing over the sphaleron barrier.
This observation has two important implications for
baryogenesis:
one is that the existence of processes of rapid baryon
number violation will invalidate the GUT baryogenesis with B-L
invariance.  Another is the
possibility of producing and maintaining excess baryon number at the
electroweak energy scale, i.e., electroweak baryogenesis (EWB).

We will focus on some of the properties of the sphaleron.
The references listed here
will be quite incomplete and we apologize for all the omissions.
 \b
{\bf II. Some Important Facts about Sphalerons}

(II.a) Energy of the sphaleron[2][4]:~~
The gauge and Higgs profile functions of the sphaleron in the SU(2)-Higgs
theory are spherically symmetric and the energy of the
sphaleron is given by
$E_{sp} = {2M_W \over {\alpha}_W}B({\lambda \over g_2^2})
         = (5 TeV)B({\lambda \over g_2^2})$, where
$2.52 \leq B({\lambda \over g_2^2})
\leq 2.70,$
$\lambda$ is the Higgs quartic coupling constant, and
$g_2$ the SU(2) coupling constant.

(II.b) Topological charge of the sphaleron[2]:~~
The sphaleron has a half-odd integer topological charge. This allows the
interpretation
that it lies on the top of the potential barrier between the vacua with
Chern-Simon numbers n and n + 1.

(II.c) Sphaleron of finite Weinberg angle[2][5]:~~ Upon
including the U(1), the sphaleron is no longer spherically
symmetric.
However, the
energy of the sphaleron differs very little from the SU(2)-Higgs
theory, slightly lower by about 1\% at the physical Weinberg angle.

(II.d) Deformed sphaleron[6]:~~
Multiple solutions due to bifurcation arise when the Higgs mass becomes
large.  The first bifurcation takes place with the appearance of new
solutions for $M_H$ $\geq$ $12m_W$. More solutions appear when the
Higgs mass increases futher.
The first bisphaleron which lies below the sphaleron has the lowest
energy, which is about 8\% lower asymptotically.

(II.e) Effect of fermions[7]:~~
The effect of the back action of the fermion
on the sphaleron energy is generally
small for light fermions less than 300 $GeV$.  For fermions as heavy as
1 $TeV$, the fermion effect on the energy of sphaleron is still less than
10\%.

(II.f) Sphaleron at finite temperature[8]:~~
The sphaleron energy can be approximated by
$E_{sp}(\lambda,T) \simeq E_{sp}(\lambda,0){< \Phi(T) >
                      \over < \Phi(0) >},$
where $< \Phi(T) >$ is the Higgs vacuum expectation value at the
temperature T.

(II.g) Strong sphaleron[9]:~~
The analysis of the Standard EW model above the symmetry restoration
temperature leads to a similar study of axial baryon number
violation in QCD at finite temperature.
There exists a sphaleron-like object
in QCD,
which has an important implication in EWB.

(II.h) Sphalerons in extended models:~~
The study of sphalerons in extended models,
one-doublet with a singlet[10] and two-doublet Higgs models[11][8],
 is motivated by the difficulties
of the one-Higgs doublet SM
with EWB: one is that the CP violation
is to too small to produce the observed
density ratio of baryons to photons. Another is that preserving
the excess baryons produced during the EW phase
transition
 requirs[12] $M_H < 45 GeV$, which
is below the current LEP experimental lower bound of $67 GeV$.
In general, the energy of the sphaleron is remarkably stable
against the variation of models of the Higgs sector.
\b
{\bf III. Beyond the standard model}

The approach we used to describe physics beyond
the SM is to add high dimension effective operators to the SM.  To
simplify the matter we take the limit of vanishing Weinberg angle and
ignore the fermion.

(III.a) Dimension 6 operators[13][14]:~~
In the absence of fermion fields, the lowest dimension is 6.  There
are the following operators:
%
%
%
${1 \over 3{\Lambda}^2}({\Phi}^{\dagger}\Phi - {v^2 \over 2})^3$;
${1 \over {\Lambda}^2}{| {\partial}_{\mu}({\Phi}^{\dagger}\Phi) |}^2$;
${1 \over {\Lambda}^2}({\Phi}^{\dagger}\Phi)(D_{\mu}\Phi)^{\dagger}
(D^{\mu}\Phi)$;
${1 \over {\Lambda}^2}{| {\Phi}^{\dagger}D_{\mu}\Phi |}^2$;
${1 \over 2{\Lambda}^2}({\Phi}^{\dagger}\Phi)W^a_{\mu\nu}W^{a\mu\nu}$.
%
%
To compute the energy of the sphaleron
we add the above operators individually to the SM
Lagrangian, and recalculate the energy.
  The ansatz of the SM profile functions is still applicable. The
contribution to the sphaleron energy depends on the value of $\Lambda$,
which we took to be 1 $TeV$. All contributions are small, to within a
few per cent of that of the SM sphaleron, except for the first operator
for small values of $\lambda$.

The first operator is anomalous for small values of $\lambda$, where the
sphaleron energy becomes very large and negative,
whether one takes the negative or positive sign of the operator.  This
change of
behavior can be understood from the fact that the
sphaleron, being classical, can probe only a limited region of the scalar
potential, i.e.,
${| \Phi |} \leq {v^2 \over 2}$.  When the potential in this
region is not affect significantly, such is the case of not too small
$\lambda$, the sphaleron energy will not suffer
much change.  For sufficiently small $\lambda$, which corresponds
to $M_H \leq 20 GeV$, the sphaleron energy can be modified drastically.
But this case is unphysical, below the experimental lower bound of the
Higgs mass.


To conclude, the inclusion of a {\it reasonably behaved}
dimension 6 operators will not modify the sphaleron energy significantly
for a reasonable cutoff like 1 $TeV$.
Some of the effects of including dimension 6 operators in the
SM scalar potential is to raise the upper limit on the Higgs mass from 45
$GeV$ to about 100 $GeV$[13].

(III.b) Dimension 8 operators[15]:~~
There are many dimension 8 operators involving the gauge and Higgs fields.
We found an interesting one in the form,
${1 \over {\Lambda}^4}\{(D_{\mu}\Phi)^{\dagger}(D^{\mu}\Phi)\}^2$.
For large $\lambda$ the sphaleron energy is proportional to
$E_{sp} \sim ({\lambda \over g^2})^{1 \over 4}$,
which blows up when the Higgs mass approaches $\infty$.
We calculated the energy numerically. The result shows that
$E^{BSM}_{sp}$ is very close to $E_{sp}$ for ${\lambda \over g^2} \leq
10^3$. $E^{BSM}_{sp}$ starts to depart from $E_{sp}$ significantly
for ${\lambda \over g^2} = 10^7$ and increases more rapidly when
$\lambda$ increases further, as shown in Fig. I.

\vskip 3in

(III.c) Sphaleron in the non-linear $\sigma$ model[16]:~~
The above result has an interesting consequence:
the sphaleron energy in the non-linear $\sigma$ model is infinity.
This result may have some bearing on
EWB in a dynamically broken theory, where the Higgs sector
is realized by a non-linear $\sigma$ model.
As the sphaleron
has a large energy in the broken phase and will satisfy the Shaposhnikov[12]
criterion, the baryon asymmetry produced in the symmetric phase
and in the bubble wall will not be washed out in the broken phase.

\ce {\bf Acknowledgement}
\no The author would like to thank Xinmin Zhang
for numerous discussions and Kerry Whisnant for reading the manuscript.
This work is supported in part
by the Office of High Energy and Nuclear Physics of the U.S. Department
of Energy (Grant No. DE-FG02-94ER40817).
\b
 \ce {\bf References}

\item{[1]} G. 't Hooft, {\it Phys. Rev. Lett.} 37 (1976) 8;
{\it Phys Rev.} D14 (1976) 3432.
\item{[2]} N.S. Manton, {\it Phys. Rev.} D28 (1983) 2019;
F.R. Klinkhamer and N.S. Manton {\it phys. Rev.}
D30 (1984) 2212.
\item{[3]} V.A. Kuzmin, V.A. Rubakov, M.E. Shaposhnikov,
{\it Phys. Lett.} 155B (1985) 36;
\item{[4]}T.Akiba, H. Kikuchi and T. Yanagida, {\it Phys. Rev.} D38
(1988) 1937; L.G. Yaffe, {\it Phys. Rev.} D43 (1989) 3463; J. Kunz and
Y. Brihaye, {\it Phys. Lett.} B216 (1989) 353
\item{[5]} B. Kleihaus, J. Kunz, and Y Brihaye, {\it Phys. Lett.}
273B (1991) 100.
\item{[6]}L.G. Yaffe, {\it Phys. Rev.} D43, (1989) 3463; J. Kunz and
 Y. Brihaye, {\it Phys. Lett.} B216 (1989) 353; F.R. Klinkhamer,
{\it Phys. Lett.} B236 (1990) 187.
\item{[7]} G. Nolte and J. Kunz, {\it Phys. Rev.} D48 (1993) 5905;
A. Roberge, {\it Phys. Rev.}, D49 (1994) R1698.
\item{[8]} M. Dine, P. Huet, and R. Singleton, {\it Nucl. Phys.} B375
(1992) 625.
S. Braibant, Y. Brihaye, and J. Kunz, {\it Int.
J. Mod. Phys.} 8 (1993) 5563.
\item{[9]} L. McLerran, E. Mottola, M.E. Shaposhnikov, {\it Phys.
Rev.} D43 (1991) 2027;
R.N. Mohapatra and Xinmin Zhang, {\it Phys. Rev.}
D45 (1992) 2699.

\item{[10]} B. Kastening and Xinmin Zhang, {\it Phys. Rev.}
D45 (1992) 3884;
K. Enqvist and I. Vilja, {\it Phys. Lett.} 287B (1992) 119.
\item{[11]} B. Kastening, R.D. Peccei, Xinmin Zhang,
{\it Phys. Lett.} 266B (1991) 413;
A.I. Bochkarev, S.V. Kuzmin, M.E. Shaposhnikov,
{\it Phys. Lett.} 244B (1990) 275; {\it Phys. Rev.} D43 (1991) 369;
N. Turok and J. Zadrozny, {\it Phys. Rev. Lett.}
65 (1990) 2331;
L. McLerran, M.E. Shaposhnikov, N. Turok and M. Voloshin,
{\it Phys. Lett.} 256B (1992) 451.

\item{[12]} M.E. Shaposhnikov, {\it Phys. Lett.} 277B (1992) 324;
{\it Nucl. Phys.} B287 (1987) 757; {\it JETP Lett.} 44 (1986) 465.

\item{[13]} Xinmin Zhang, {\it Phys. Rev.} D47 (1993) 3065.
\item{[14]}Xinmin Zhang and Bing-Lin Young, {\it Phys. Rev.} D49
(1994) 563;
Seungkoog Lee. J. Spence, and Bing-Lin Young, Iowa State
preprint.
\item{[15]} Xinmin Zhang, Seungkoog Lee, and Bing-Lin Young,
to be published in Phys. Rev. D (hep-ph/9406322).
\item{[16]} F.R. Klinkhamer and J. Boguta, {\it Z. Phys.}
C40 (1988) 415.

\bye
\bye